\documentclass[pra,showpacs,floatfix]{revtex4}
\usepackage{amssymb}
\usepackage{graphicx}

\begin{document}

\title{Interface solitons in one-dimensional locally-coupled lattice systems}
\author{ Lj. Had\v zievski$^1$, G. Gligori\'c$^1$, A. Maluckov$^2$, and B.
A. Malomed$^3$}
\affiliation{$^1$ Vin\v ca Institute of Nuclear Sciences, P.O. Box 522,11001 Belgrade,
Serbia \\
$^2$ Faculty of Sciences and Mathematics, P.O.B. 224, 18000 Ni\v s, Serbia\\
$^3$ Department of Physical Electronics, School of Electrical Engineering,
Faculty of Engineering, Tel Aviv University, Tel Aviv 69978, Israel}

\begin{abstract}
Fundamental solitons pinned to the interface between two discrete lattices
coupled at a single site are investigated. Serially and parallel-coupled
identical chains (\textit{System 1} and \textit{System 2}), with the
self-attractive on-site cubic nonlinearity, are considered in one dimension.
In these two systems, which can be readily implemented as arrays of
nonlinear optical waveguides, symmetric, antisymmetric and asymmetric
solitons are investigated by means of the variational approximation (VA) and
numerical methods. The VA demonstrates that the antisymmetric solitons exist
in the entire parameter space, while the symmetric and asymmetric modes can
be found below some critical value of the coupling parameter. Numerical
results confirm these predictions for the symmetric and asymmetric
fundamental modes. The existence region of numerically found antisymmetric
solitons is also limited by a certain value of the coupling parameter. The
symmetric solitons are destabilized via a supercritical symmetry-breaking
pitchfork bifurcation, which gives rise to stable asymmetric solitons, in
both systems. The antisymmetric fundamental solitons, which may be stable or
not, do not undergo any bifurcation. In bistability regions stable
antisymmetric solitons coexist with either symmetric or asymmetric ones.
\end{abstract}

\maketitle

\section{Introduction}

The study of surface modes in multi-layered optical media has begun long ago
\cite{early}. Recently, a great deal of attention was attracted to studies
of solitons pinned to interfaces between different nonlinear optical media,
at least one of which carries a lattice structure. Such \textit{surface
solitons} were predicted theoretically in diverse settings \cite%
{surface-theory} and soon after that created in experiments \cite%
{surface-experiment}. Surface solitons were also considered in the general
context of junctions between different discrete lattices \cite{3}. In these
studies, it was concluded that the self-trapped surface modes acquire novel
properties, different from those of the solitons known in uniform lattices.
In particular, discrete surface states can only exist above a certain
threshold value of the total power (the soliton's norm), and a bistability
is possible, with different surface modes coexisting at a common value of
the power.

Generally speaking, these nonlinear modes may be considered as a variety of
optical solitons pinned by defects, which were also studied theoretically
\cite{defect-theory,molinakivshar} and experimentally \cite%
{defect-experiment,Roberto} in many systems, continual and discrete (a
comprehensive review of the topic of discrete and lattice solitons in
optics, including the interaction with defects, was given in recent articles
\cite{big-review}; for a general review of discrete solitons, see book \cite%
{PGK}). These studies have demonstrated that discrete nonlinear photonic
systems may support spatially localized states with different symmetries,
which can be controlled by the insertion of suitable defects into the
lattice \cite{molinakivshar}. Localized modes supported by defects of
optical lattices (OLs)\ were also studied in models of Bose-Einstein
condensates (BECs) \cite{Konotop}.

Another topic which is relevant to the present work is the possibility of
the \textit{spontaneous symmetry breaking} (SSB) in two-mode symmetric
settings with a linear coupling between two subsystems. An alternative
realization of settings which give rise to the SSB is represented by
double-well potentials (in that case, the two coupled modes correspond to
states trapped in the two symmetric wells). A specific version of the SSB
corresponds to the double-well \textit{pseudopotential}, induced by a
symmetric spatial modulation of the local nonlinearity coefficient \cite%
{Dong}.

SSB bifurcations, which destabilize symmetric states and give rise to
asymmetric ones, were originally predicted in terms of the self-trapping in
discrete systems \cite{Scott}. In the physically important model of
dual-core nonlinear fibers, the SSB instability was discovered in Ref. \cite%
{Wright}, and the respective bifurcations for continuous-wave states
were studied in detail in Ref. \cite{Snyder}, for various types of
the intra-core nonlinearities. It is also relevant to mention early
work \cite{earlier}, which put forward the SSB\ concept in the
framework of the nonlinear Schr\"{o}dinger (NLS) equation. Further,
the SSB was studied in detail for solitons in the model of the
dual-core fiber with the cubic (Kerr) nonlinearity
\cite{Akhmediev,malomed}. Similar analysis was later performed
for gap solitons in the models of dual-core \cite{Mak} and tri-core \cite%
{gubeskys1} fiber Bragg gratings, and for matter-wave solitons in
the BEC loaded into a dual-core potential trap combined with a
longitudinal OL. The latter analysis was performed in the models
with both one \cite{1D-BEC} and two \cite{2D-BEC} longitudinal
dimensions (1D and 2D, respectively).

The limit case of a very strong OL corresponds to a discrete lattice \cite%
{wannier}. In that case, the SSB of 1D and 2D discrete solitons in the
system of two linearly coupled discrete NLS equations (DNLSEs) was
investigated in Ref. \cite{VA}.

In addition to the studies of the SSB in diverse two-core systems with the
cubic nonlinear terms, this effect was also analyzed in models describing
optical media with quadratic \cite{chi2} and cubic-quintic \cite{CQ}
nonlinearities. In the latter case, the SSB diagrams feature loops, with
asymmetric solitons existing at intermediate values of the total power,
while only symmetric modes can be found at low and high powers.

General conclusions about the character of the SSB in solitons can
be drawn from the above-mentioned works. The self-focusing
nonlinearity induces the SSB of symmetric modes, with a trend to
make the respective bifurcation \textit{subcritical}, whose
characteristic feature is a bistability region, in which stable
symmetric and asymmetric solitons coexist. The self-focusing
nonlinearity does not induce any bifurcation of antisymmetric
solitons. In the absence of the periodic potential induced by the
OL, the entire family of the antisymmetric solitons is unstable.

However, if the self-attractive nonlinearity acts in the combination with a
sufficiently strong OL, the antisymmetric solitons may be stable. In the
same case, the SSB bifurcation is transformed from subcritical into the
\textit{supercritical} one, which does not admit the coexistence of stable
symmetric and asymmetric solitons. Nevertheless, a global bistability of a
different type takes place in this situation, as stable antisymmetric
solitons coexist with their symmetric and asymmetric counterparts, below and
above the bifurcation point, respectively (in the 2D setting, both families
of the asymmetric and antisymmetric solitons attain a termination point at
high values of the total power, due to the onset of the collapse driven by
the self-attraction).

On the other hand, in the models combining the self-defocusing
nonlinearity and OL potentials, symmetric solitons do not suffer
any SSB,
while antisymmetric solitons are destabilized by the \textit{%
anti-symmetry-breaking} pitchfork bifurcation, which is always of the
supercritical type.

In this work we aim to study localized modes at the interface of two 1D
uniform discrete lattices (chains) with the cubic onsite nonlinearity, which
are linearly coupled either in series (\textit{System 1}, see Fig. \ref%
{fig1} below) or in parallel (\textit{System 2}, see Fig. \ref{fig4}%
). In fact, the former configuration may be realized as a single
discrete lattice with a \textit{spring defect}, in the form of a
locally modified inter-site coupling constant (such a system has
been actually created in optics, as an array of nonlinear
waveguides with a modified separation between two of them
\cite{Roberto}). System 2 differs from that introduced in Ref.
\cite{VA} by the fact that the two uniform identical chains placed
in parallel planes are linearly coupled in the transverse
direction at a singe site, rather than featuring the uniform
linear coupling. Analyzing localized defect/surface modes in these
two systems, we focus on their symmetry properties -- in
particular, with the intention to investigate SSB transitions in
them.

The presentation is organized as follows. The systems, 1 and 2, are
introduced in Section 2, where we consider the existence and stability of
various localized modes, applying the variational approximation (VA) and the
Vakhitov-Kolokolov (VK) stability criterion. Dynamical properties of the
discrete solitons with different symmetries are investigated by means of
numerical methods, including the computation of the linear-stability
eigenvalues and direct simulations in Section 4. The paper is concluded by
Section 5.

\section{Fundamental surface solitons in the system of two coupled lattices}

\subsection{System 1: the serial coupling}

\begin{figure}[h]
\center\includegraphics [width=8cm]{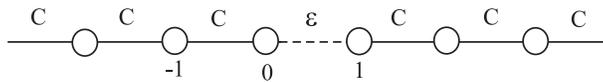} \caption{A schematic
of System 1, which consists of two semi-infinite
identical chains, linked by the modified linear coupling $\protect%
\varepsilon $ between sites $n=0$ and $n=1$.} \label{fig1}
\end{figure}

The system formed by two linked identical semi-infinite chains is
displayed in Fig. \ref{fig1}, with inter-site coupling constant
$C$ inside the chains, and constant $\varepsilon $ accounting for
the linkage between them. We consider only the case when $C$ and
$\varepsilon $ have the same sign, as opposite signs of the
inter-site couplings are difficult to realize in optical and BEC
systems. The on-site nonlinearity is cubic, with the respective
coefficient, $\gamma $, being constant throughout the system.
Thus, this model is described by the following DNLSE system,%
\begin{eqnarray}
i\frac{d\phi _{n}}{dz}+\phi _{n+1}+\phi _{n-1}+\left\vert \phi
_{n}\right\vert ^{2}\phi _{n} &=&0,\ n\neq 0,1,  \nonumber \\
i\frac{d\phi _{0}}{dz}+\varepsilon \phi _{1}+\phi _{-1}+\left\vert \phi
_{0}\right\vert ^{2}\phi _{0} &=&0,  \nonumber \\
i\frac{d\phi _{1}}{dz}+\varepsilon \phi _{0}+\phi _{2}+\left\vert \phi
_{1}\right\vert ^{2}\phi _{1} &=&0,  \label{eq3a}
\end{eqnarray}%
where $z$ is the propagation distance (assuming that the chains
correspond to two semi-infinite arrays of parallel optical
waveguides), and an obvious rescaling is used to fix $C=\gamma =1$
[that is, $\varepsilon $ which
figures in Eqs. (\ref{eq3a}) is $\varepsilon /C$, in terms of Fig. \ref%
{fig1}].

This system may be considered as the usual DNLSE containing a local "spring
defect". As mentioned above, such a system has been created in optics, in
the form of a regular array of parallel waveguiding cores, with a change of
the distance between two of them \cite{Roberto}. The interaction of lattice
solitons with this array defect was studied experimentally in Ref. \cite%
{Roberto}.

Stationary solutions for solitons formed at the interface between the chains
are looked for as usual, $\phi _{n}=u_{n}\exp \left( i\mu z\right) $, where $%
u_{n}$ and $\mu $ are real lattice field and propagation constant. The
corresponding stationary equations following from Eqs. (\ref{eq3a}) are
\begin{eqnarray}
-\mu u_{n}+u_{n+1}+u_{n-1}+u_{n}^{3} &=&0,\ n\neq 0,1,  \nonumber \\
-\mu u_{0}+\varepsilon u_{1}+u_{-1}+u_{0}^{3} &=&0,  \nonumber \\
-\mu u_{1}+\varepsilon u_{0}+u_{2}+u_{1}^{3} &=&0.  \label{eq3}
\end{eqnarray}

\subsubsection{The variational approximation}

Equations (\ref{eq3}) can be derived from the Lagrangian,%
\begin{equation}
L=L_{-}+L_{+}+2\varepsilon u_{0}u_{1},  \label{eq6}
\end{equation}%
\begin{eqnarray*}
L_{-} &\equiv &\sum_{n=-\infty }^{-1}\left( -\mu u_{n}^{2}+\frac{1}{2}%
u_{n}^{4}+2u_{n}u_{n+1}\right) +\left( -\mu u_{0}^{2}+\frac{1}{2}%
u_{0}^{4}\right) ,~ \\
L_{+} &\equiv &\sum_{n=1}^{\infty }\left( -\mu u_{n}^{2}+\frac{1}{2}%
u_{n}^{4}+2u_{n}u_{n+1}\right) ,
\end{eqnarray*}%
where $L_{\pm }$ are the intrinsic Lagrangians of the two semi-infinite
chains, and the last term in Eq. (\ref{eq6}) accounts for the coupling
between them. To apply the variational approximation (VA), we follow Ref.
\cite{VA} and adopt an \textit{ansatz} consisting of two parts:
\begin{eqnarray}
u_{n} &=&A\exp {(an)},\ \mathrm{for}\ n\leq 0,  \nonumber \\
u_{n} &=&B\exp {(-a(n-1))},\ \mathrm{for}\ n\geq 1.  \label{eq7}
\end{eqnarray}%
This ansatz admits different amplitudes, $A\neq B$, of the solution in the
linked chains, but postulates a common width of the ansatz in both of them, $%
a^{-1}$. Actually, the SSB is represented by the appearance of asymmetric
solutions, with $A^{2}\neq B^{2}$.

Amplitudes $A$ and $\,B$ are treated below as variational parameters. As
concerns inverse width $a$, it is fixed through a solution of the
linearization of Eqs. (\ref{eq3}) at $|n|\rightarrow \infty $,%
\begin{equation}
a=\ln \left( \mu /2+\sqrt{\mu ^{2}/4-1}\right) ,  \label{a}
\end{equation}%
which implies that the propagation constant takes values $\mu >2$. Relation (%
\ref{a}) may also be represented in the following forms, that will be used
below:
\begin{equation}
s\equiv e^{-a}=\frac{\mu }{2}-\sqrt{\frac{\mu ^{2}}{4}-1},~\mu =s+s^{-1}.
\label{s}
\end{equation}

The substitution of ansatz (\ref{eq7}) into Eq. (\ref{eq6}) yields the
corresponding effective Lagrangian, where Eq. (\ref{a}) is used to eliminate
$\mu $ in favor of $s$:
\begin{equation}
L_{\mathrm{eff}}=\left( L_{-}\right) _{\mathrm{eff}}+\left( L_{+}\right) _{%
\mathrm{eff}}+2\varepsilon AB,  \label{eq8}
\end{equation}%
\begin{eqnarray}
\left( L_{-}\right) _{\mathrm{eff}} &=&-s^{-1}A^{2}+\frac{1}{2\left(
1-s^{4}\right) }A^{4}  \nonumber \\
\left( L_{+}\right) _{\mathrm{eff}} &=&-s^{-1}B^{2}+\frac{1}{2\left(
1-s^{4}\right) }B^{4}.  \label{eq10}
\end{eqnarray}%
This Lagrangian gives rise to the Euler-Lagrange equations for amplitudes $A$
and $B$: $\left( \partial /\partial A\right) \left( L_{-}\right) _{\mathrm{%
eff}}+2\varepsilon B=0,\left( \partial /\partial B\right) \left(
L_{+}\right) _{\mathrm{eff}}+2\varepsilon A=0$, or, in the explicit form,%
\begin{eqnarray}
-s^{-1}A+\frac{1}{1-s^{4}}A^{3}+\varepsilon B &=&0  \nonumber \\
-s^{-1}B+\frac{1}{1-s^{4}}B^{3}+\varepsilon A &=&0.  \label{eq12}
\end{eqnarray}%
These equations allow us to predict the existence of three different species
of the localized modes: symmetric ones, with $A=B$, antisymmetric with $A=-B$%
, and asymmetric with $A^{2}\neq B^{2}$.

The results of the VA are reported below, along with the respective
numerical results. It will be seen that the VA predicts characteristics of
the symmetric and asymmetric modes very accurately, while the approximation
is essentially less accurate for the antisymmetric solitons.

\subsubsection{Existence regions for the interface solitons}

\begin{figure}[h]
\center\includegraphics [width=13cm]{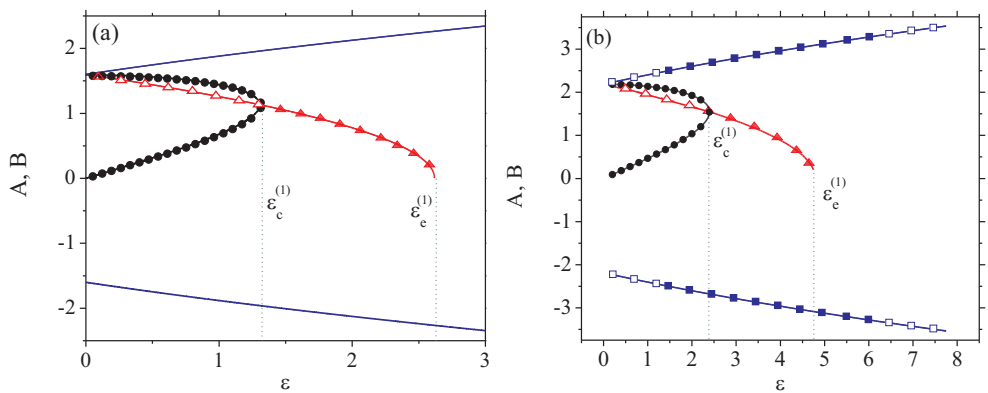} \caption{(Color
online) Amplitudes $A$ and $B$ of asymmetric, symmetric, and
antisymmetric solitons (black, red, and blue colors,
respectively), as predicted by the VA (lines) and obtained in the
numerical form (AS - circles, SyS - triangles, AnS - squares),
versus $\protect\varepsilon $, for System 1, at fixed values of
the propagation constant: $\protect\mu =3$ (a) and $\protect\mu
=5$ (b). The
dotted green vertical lines denote the critical values of $\protect%
\varepsilon $ limiting the existence regions of the asymmetric ($\protect%
\varepsilon _{c}$) and symmetric ($\protect\varepsilon _{e}$) solitons. For $%
\protect\mu =3$, antisymmetric solitons were not found in the numerical
form. Empty and full symbols pertain to unstable and stable solitons,
respectively, as concluded from the numerical investigation.}
\label{fig2}
\end{figure}

\begin{figure}[h]
\center\includegraphics [width=13cm]{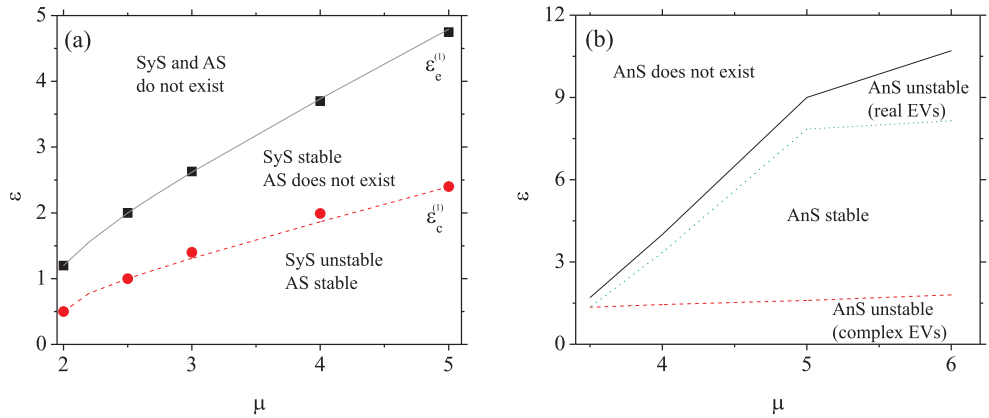} \caption{(Color
online) (a) The existence and stability diagrams for the
fundamental symmetric (SyS) and asymmetric (AS) solitons in System
1, generated by means of the VA (curves) and numerical
computations (symbols), in the plane of $\left(
\protect\varepsilon ,\protect\mu \right) $. Black squares, which
mark the boundary of the SyS existence region, are located
very close to the respective line $\protect\varepsilon _{e}^{(1)}(\protect%
\mu )$ predicted by the VA (see Fig. \protect\ref{fig2}). Red
circles, which denote the numerically found boundary of the AS
existence region, are also close to the VA-predicted curve
$\protect\varepsilon _{c}^{(1)}(\protect\mu ) $, along which the
symmetry-breaking bifurcation takes place, destabilizing the SyS
and giving birth to two mutually symmetric branches of the AS
modes. (b) The existence and stability borders for the
antisymmetric mode (AnS) in the same parameter plane, found in the
numerical form only (the variational approximation does not
predict existence limits for the AnS). Inside the instability
areas in (b), the type of the respective unstable eigenvalues
(EVs) of small perturbations is indicated.} \label{fig3}
\end{figure}

The solution for the symmetric solitons (SyS) is easily obtained from Eqs. (%
\ref{eq12}),
\begin{equation}
A^{2}=({1-s^{4}})(s^{-1}-\varepsilon ).  \label{eq13}
\end{equation}%
The dependence of this amplitude on $\varepsilon $ for fixed $s=\left( 5-%
\sqrt{21}\right) /2\approx \allowbreak 0.21$, which corresponds to
$\mu =5$, is plotted in Fig. \ref{fig2}. As follows from Eq.
(\ref{eq13}), in System 1 the existence domain of the SyS
solutions at a given value of the propagation constant, $\mu $, is
limited to $\varepsilon <${$\varepsilon _{e}^{(1)}$}$\equiv
s^{-1}$. The upper existence limit for the SyS is seen in Fig.
\ref{fig2}, and also in Fig. \ref{fig3}(a), which shows the
existence regions predicted by the VA for all the types of the
solitons in the plane of $\left( \varepsilon ,\mu \right) $, along
with the numerically found counterparts of these regions (the
procedure of obtaining numerical results is described below).

The existence range of the asymmetric solution (AS), with $A^{2}\neq B^{2}$,
can be estimated by taking the difference of two equations (\ref{eq12}) and
canceling a common factor $\left( A-B\right) $:
\begin{equation}
\left( s^{-1}+\varepsilon \right) =A^{2}+B^{2}+AB.  \label{eq14}
\end{equation}%
In the limit of $A-B\rightarrow 0$, Eq. (\ref{eq14}) yields the following
relation:
\begin{equation}
A^{2}=(1/3)({1-s^{4}})\left( s^{-1}-\varepsilon \right) .  \label{eq15}
\end{equation}%
Equating expressions (\ref{eq13}) and (\ref{eq15}), one can predict the
value of the linkage constant at which the \textit{SSB bifurcation} gives
birth to the pair of AS in System 1,
\begin{equation}
{\varepsilon }_{c}^{(1)}=\left( 2s\right) ^{-1}.  \label{eq16}
\end{equation}%
For broad solitons, with $a\rightarrow \infty $, i.e., $s\rightarrow 1$ [see
Eq. (\ref{eq7})], {$\varepsilon $}$_{c}^{(1)}$ given by Eq. (\ref{eq16})
approaches a limit value, {$\varepsilon $}$_{c}^{(1)}=1/2$, while for
strongly localized solitons, with $\mu \rightarrow \infty $ and $s<<1$,
relation (\ref{s}) demonstrates that Eq. (\ref{eq16}) yields {$\varepsilon $}%
$_{c}^{(1)}\approx \mu /2$, cf. Figs. \ref{fig2} and \ref{fig3}).

The general solution for the AS can be found after some algebraic
manipulations with Eqs. (\ref{eq12}) (replacing the equations by their sum
and difference, and solving the latter equations for $A^{2}+B^{2}$ and $AB$,
cf. similar exact solutions obtained for the SSB\ in the double-well
pseudopotential in Ref. \cite{Dong}):
\begin{eqnarray}
A &=&\frac{\varepsilon \sqrt{2s({1-s^{4}})}}{\sqrt{1+\sqrt{1-4s^{2}{%
\varepsilon }^{2}}}},  \nonumber \\
B &=&\sqrt{\frac{{1-s^{4}}}{2s}}\sqrt{1+\sqrt{1-4s^{2}{\varepsilon }^{2}}}.
\label{eq20}
\end{eqnarray}%
This solution exists at ${\varepsilon <\varepsilon }_{c}$, where ${%
\varepsilon }_{c}$ is precisely the bifurcation value given by Eq. (\ref%
{eq16}). The conclusion that the asymmetric modes exist when the coupling
constant \ ($\varepsilon $) is not too large is very natural \cite{Akhmediev}%
-\cite{2D-BEC}, \cite{VA}. Indeed, the extreme case of the asymmetric
solution, with $A\neq 0$ and $B=0$, is obviously possible in the limit of $%
\varepsilon =0$. Equally natural is the conclusion that the
symmetric mode is stable at ${\varepsilon >\varepsilon }_{c}$ and
unstable at ${\varepsilon <\varepsilon }_{c}$, while the
asymmetric mode is stable in the entire region of its existence,
as seen in Fig. \ref{fig2}. Finally, we notice that the SSB\
bifurcations observed in Fig. \ref{fig2} (and in Fig. \ref{fig5}
below, which pertains to System 2) are clearly the pitchfork
bifurcations of the \textit{supercritical} type, on the contrary
to the \textit{subcritical} bifurcation reported in Ref. \cite{VA}
for the model based on two uniformly coupled parallel chains.

For antisymmetric solitons (AnS) the amplitude is obtained from Eq. (\ref%
{eq12}) by substituting $A=-B$, which yields
\begin{equation}
A^{2}=\left( {1-s^{4}}\right) \left( s^{-1}+\varepsilon \right) .
\label{eq17}
\end{equation}%
Relation (\ref{eq17}) is plotted versus $\varepsilon $ for fixed
$\mu =5$ by blue lines in Fig. \ref{fig2}. As follows from this
relation, the VA predicts the existence of the AnS in the entire
parameter space, on the contrary to the limited existence regions
predicted for the SyS and AS modes. However, numerical results
reported below demonstrate that the AnS also have their existence
limits, see Fig. \ref{fig3}(b).

\subsubsection{Stability}

The stability of the discrete solitons predicted by the VA can be first
estimated by dint of the VK criterion, $dP/d\mu >0$, where $P\equiv
\sum_{n=-\infty }^{n=+\infty }u_{n}^{2}$ is the total power (norm) of the
soliton \cite{isrl}. Being necessary, but not the sufficient, this condition
should be combined with the spectral criterion, which requires all the
eigenvalues of small perturbations around the solitons to be stable.

The power corresponding to ansatz (\ref{eq7}) is
\begin{equation}
P=\left( 1-s^{2}\right) ^{-1}({A^{2}}+B^{2}).  \label{eq21}
\end{equation}%
For the SyS solution with $A=B,$ Eqs. (\ref{eq21}) and (\ref{eq13}) yield $%
P=2(1+s^{2})\left( s^{-1}-\varepsilon \right) $. This expression satisfies
condition $\partial P/\partial s<0$, which is tantamount to the VK
criterion, at all values of $\varepsilon $. \ Nevertheless, it is obvious
that the the branch of the symmetric solitons is destabilized by the SSB
bifurcation, hence it is unstable at $\varepsilon <\varepsilon _{c}$, as
shown in Fig. \ref{fig2} (it is well known that the respective instability
is not detected by the VK criterion \cite{Akhmediev}-\cite{2D-BEC}, \cite{VA}%
).

For the AS solutions, the substitution of Eq. (\ref{eq20}) into Eq. (\ref%
{eq21}) produces a simple expression for the total power, which does not
depend on $\varepsilon $:
\begin{equation}
P=\left( 1+s^{2}\right) /s.  \label{independent}
\end{equation}%
It also satisfies the VK criterion in the entire region of the existence of
the asymmetric interface modes.

Finally, for the antisymmetric modes in System 1, with $A=-B$, the use of
Eq. (\ref{eq17}) gives $P=2(1+s^{2})\left( s^{-1}+\varepsilon \right) $. In
this case, condition $dP/ds<0$ is satisfied \emph{only} at
\begin{equation}
\varepsilon >{\varepsilon }_{s}^{(1)}=\left( 1-s^{2}\right) s^{-3},
\label{eq28}
\end{equation}%
while the antisymmetric modes are \textit{VK-unstable} in the other part of
their existence region, at $\varepsilon <\varepsilon _{s}$. This is in
contrast with the above conclusions that the symmetric and asymmetric
solitons comply with the VK criterion in their entire existence domains.

\subsubsection{Comparison to the uniform chain}

It is instructive to compare the results predicted for System 1 by the VA to
the well-known properties of the uniform infinite chain described by the
usual DNLSE, which corresponds to Eqs. (\ref{eq3a}) with $\varepsilon =1$
\cite{PGK}. To this end, one may consider cuts of Figs. \ref{fig2}(a) and
(b) through $\varepsilon =1$. Along these cuts, one finds an unstable SyS,
stable AS, and stable AnS. In terms of the infinite uniform chain, the SyS
corresponds to an \textit{inter-site-centered} (alias \textit{off-site})
discrete soliton, which is, indeed, always unstable in the usual DNLSE \cite%
{PGK}. The fact that the \textit{spring defect} with $\varepsilon >1$ makes
the off-site soliton stable beyond the bifurcation point, i.e., at $%
\varepsilon <\varepsilon _{c}$, is remarkable by itself.

Further, what we define as the AS, corresponds to the usual on-site-centered
soliton in the ordinary DNLSE, which is always stable, as is well known. And
finally, the AnS soliton is tantamount to the localized \textit{twisted mode}
in the DNLSE setting \cite{twisted}, which is stable with small amplitudes
and unstable if its amplitude is larger, as in the case corresponding to
Fig. \ref{fig2}(b) at $\varepsilon =1$.

\subsection{System 2: the parallel coupling}

\begin{figure}[h]
\center\includegraphics [width=8cm]{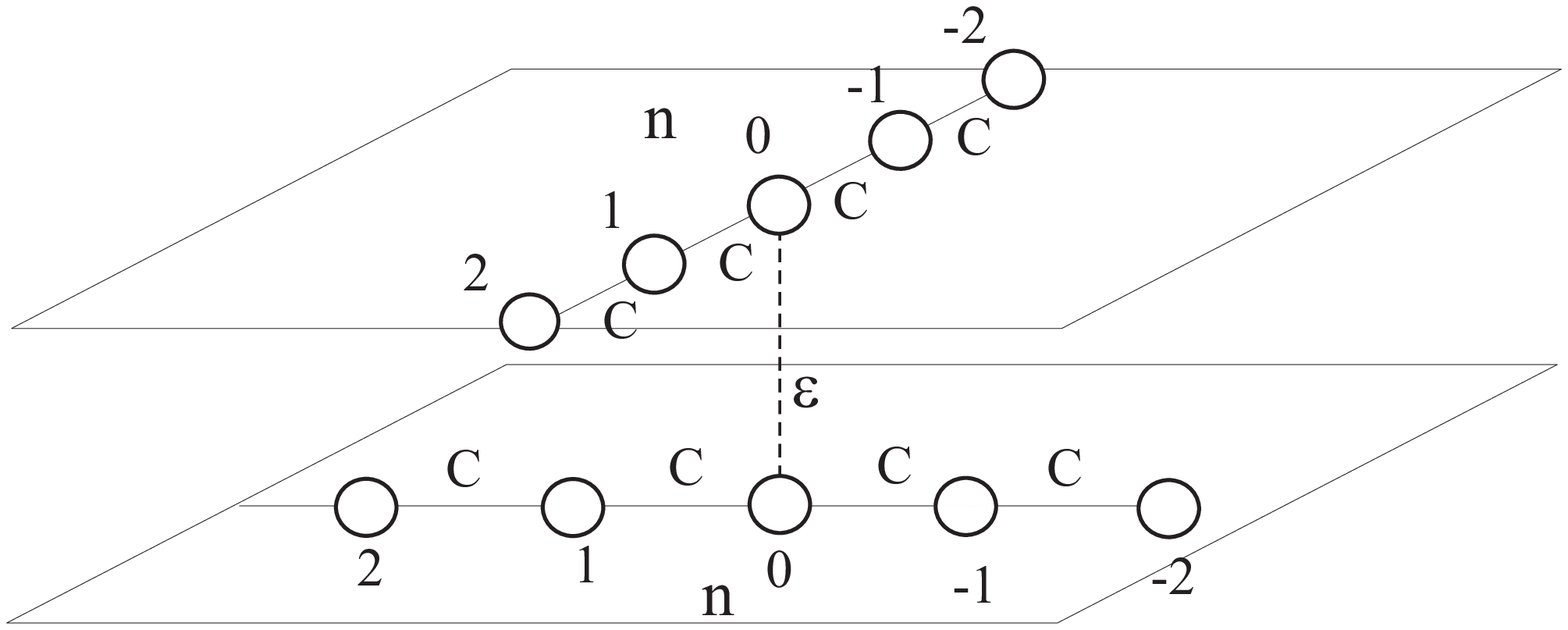} \caption{The
schematic of System 2, which is built of two uniform identical
chains placed in parallel planes, which are linearly coupled in
the transverse direction at site n = 0. The longitudinal and
transverse coupling constants are C and $\varepsilon$,
respectively.} \label{fig4}
\end{figure}

The system formed by two identical infinite uniform chains,
coupled by the transverse link at the single site, is shown in
Fig. \ref{fig4}.

The corresponding DNLSE system is (where, as well as in Eqs. (\ref{eq3a}),
we set $C=\gamma =1$, by means of rescaling):
\begin{eqnarray}
i\frac{d\phi _{n}}{dz}+\phi _{n+1}+\phi _{n-1}\gamma \left\vert \phi
_{n}\right\vert ^{2}\phi _{n} &=&0,\ n\neq 0,  \nonumber \\
i\frac{d\phi _{0}}{dz}+\phi _{1}+\phi _{-1}+\varepsilon \psi _{0}+\left\vert
\phi _{0}\right\vert ^{2}\phi _{0} &=&0,  \nonumber \\
i\frac{d\psi _{n}}{dz}+\psi _{n+1}+\psi _{n-1}+\left\vert \psi
_{n}\right\vert ^{2}\psi _{n} &=&0,\ n\neq 0,  \nonumber \\
i\frac{d\psi _{0}}{dz}+\psi _{1}+\psi _{-1}+\varepsilon \phi _{0}+\gamma
\left\vert \psi _{0}\right\vert ^{2}\psi _{0} &=&0.  \label{eq1c2}
\end{eqnarray}%
The stationary solutions are looked for in the form of $\phi
_{n}=u_{n}\exp \left( i\mu z\right) $ and $\psi _{n}=v_{n}\exp
\left( i\mu z\right) ,$ with a common propagation constant, $\mu
$, and real functions $u_{n}$ and $v_{n}$ satisfying the following
equations:
\begin{eqnarray}
-\mu u_{n}+u_{n+1}+u_{n-1}+\left\vert u_{n}\right\vert ^{2}u_{n} &=&0;\
n\neq 0  \nonumber \\
-\mu v_{n}+v_{n+1}+v_{n-1}+\left\vert v_{n}\right\vert ^{2}v_{n} &=&0;\
n\neq 0  \nonumber \\
-\mu u_{0}+u_{1}+u_{-1}+\varepsilon v_{0}+\left\vert u_{0}\right\vert
^{2}u_{0} &=&0;  \nonumber \\
-\mu v_{0}+v_{1}+v_{-1}+\varepsilon u_{0}+\left\vert v_{0}\right\vert
^{2}v_{0} &=&0.  \label{eq29}
\end{eqnarray}

\subsubsection{The variational approximation}

The stationary equations can be derived from the Lagrangian, $%
L=L_{1}+L_{2}+2\varepsilon u_{0}v_{0},$where $L_{1}$ and $L_{2}$ are the
intrinsic Lagrangians of the two uniform chains. Similar to System 1, the VA
is based on the exponentially localized ansatz with possibly different
amplitudes $A$ and $B$ but a common width, $a^{-1}$, cf. Eqs. (\ref{eq7}):
\begin{equation}
u_{n}=A\exp {(-a|n|),~}v_{n}=B\exp {(-a|n|)}.  \label{eq31}
\end{equation}%
By substituting this ansatz into the Lagrangian, and using, as above, Eq. (%
\ref{s}) to eliminate $\mu $, we obtain
\begin{equation}
L_{\mathrm{eff}}=-\frac{1-s^{2}}{s}\left( A^{2}+B^{2}\right) +\frac{1}{2}%
\frac{1+s^{4}}{1-s^{4}}\left( A^{4}+B^{4}\right) +2\varepsilon AB,
\label{eq32}
\end{equation}%
cf. Eqs. (\ref{eq8}) and (\ref{eq10}). Finally, the Euler-Lagrange
equations, $\partial L_{\mathrm{eff}}/\partial A=\partial L_{\mathrm{eff}%
}/\partial B=0$, applied to Lagrangian (\ref{eq32}), take the following
form:
\begin{eqnarray}
-\frac{1-s^{2}}{s}A+\frac{1+s^{4}}{1-s^{4}}A^{3}+\varepsilon B &=&0
\nonumber \\
-\frac{1-s^{2}}{s}B+\frac{1+s^{4}}{1-s^{4}}B^{3}+\varepsilon A &=&0.
\label{eq33}
\end{eqnarray}

\subsubsection{Existence regions for soliton solutions}

\begin{figure}[h]
\center\includegraphics [width=6.5cm]{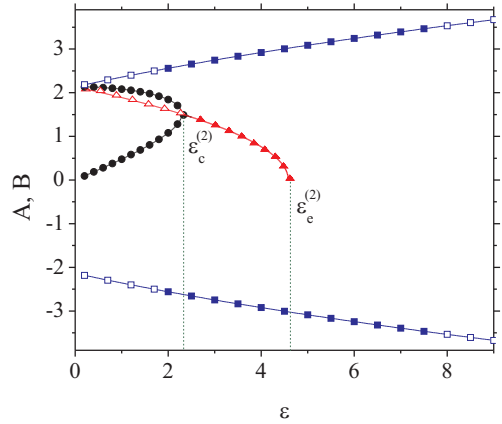} \caption{(Color
online) The same as in Fig. \protect\ref{fig2}(b) (i.e., with
$\protect\mu =5$), but for System 2.} \label{fig5}
\end{figure}

\begin{figure}[h]
\center\includegraphics [width=13cm]{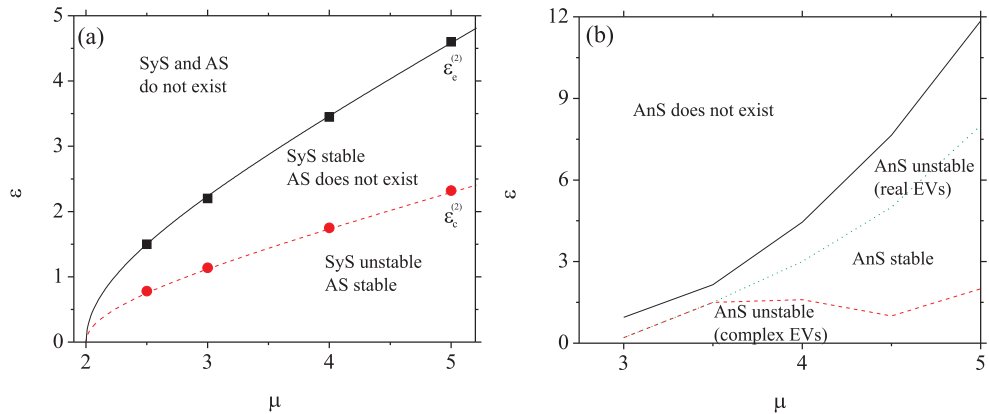} \caption{(Color
online) The same as in Fig. \protect\ref{fig3}, but for System 2.}
\label{fig6}
\end{figure}

The solution for the SyS (symmetric soliton), with~$A=B$, is immediately
obtained from Eq. (\ref{eq33}):
\begin{equation}
A^{2}=\frac{1-s^{4}}{1+s^{4}}\left( \frac{1-s^{2}}{s}-\varepsilon \right) .
\label{eq34}
\end{equation}%
This relation is plotted in Fig. \ref{fig5} for $\mu =5$. Similar
to the case of System 1, it follows from Eq. (\ref{eq34}) that in
System 2 the SyS exists at $\varepsilon <${$\varepsilon
_{e}^{(2)}$}$=\left( 1-s^{2}\right) /s $, as shown in Figs.
\ref{fig5} and \ref{fig6}. In the latter figure, the critical
value $\varepsilon _{e}^{(2)}$ is plotted by the black line.

The existence range of the AS solutions can be determined by means of the
same procedure which was used above in the case of System 1. The result is
that, in the present case, the asymmetric modes exist at
\begin{equation}
\varepsilon <{\varepsilon }_{c}^{(2)}=\left( 1-s^{2}\right) /\left(
2s\right) ,  \label{eq35}
\end{equation}%
cf. Eq. (\ref{eq16}). For the broad asymmetric solitons $(s\rightarrow 1)$, {%
$\varepsilon _{c}^{(2)}$} approaches the limiting value {$\varepsilon
_{c}^{(2)}$}$=0$, while for the strongly localized solitons $(s<<1)$ we
obtain {$\varepsilon $}$_{c}\approx \mu /2$. The full solution for the AS
mode can be found too, cf. Eq. (\ref{eq20}) for System 1:%
\begin{eqnarray}
A &=&\sqrt{\frac{{1-s^{4}}}{2s(1+s^{4})}}\sqrt{\left( 1-s^{2}\right) +\sqrt{%
(1-s^{2})^{2}-4s^{2}{\varepsilon }^{2}}}  \nonumber \\
B &=&\varepsilon\sqrt{\frac{2s(1-s^4)}{1+s^4}} \frac{1}{\sqrt{{1-s}^{2}+%
\sqrt{(1-s^{2})^{2}-4s^{2}{\varepsilon }^{2}}}}  \label{eq37}
\end{eqnarray}%
Note that the AS solution exists at $\varepsilon <\left(
1-s^{2}\right) /\left( 2s\right) $, which is exactly equivalent to
Eq. (\ref{eq35}), obtained above by means of a different algebra.
The amplitudes of AS at the interface sites are plotted in Fig.
\ref{fig5}.

The solution for the antisymmetric $(A=-B)$ solitons (AnS) produced by Eqs. (%
\ref{eq33}) is
\begin{equation}
A^{2}=\frac{1-s^{4}}{1+s^{4}}\left( \frac{1-s^{2}}{s}+\varepsilon \right) .
\label{eq36}
\end{equation}%
As seen from here, the AnS are predicted by the VA to exist in the entire
parameter space.

\subsubsection{Stability}

The VK criterion may also be applied to the soliton solutions predicted in
System 2. The total power of the solutions based on ansatz (\ref{eq31}) is
\begin{equation}
P=\frac{1+s^{2}}{1-s^{2}}({A^{2}}+B^{2}).  \label{eq38}
\end{equation}

For the SyS solution (\ref{eq34}) with $A=B$, Eq. (\ref{eq38}) yields
\begin{equation}
P=\frac{2(1+s^{2})^{2}}{1+s^{4}}\left( \frac{1-s^{2}}{s}-\varepsilon \right)
,  \label{eq39}
\end{equation}%
which satisfies the VK condition, $\partial P/\partial s<0$ (tantamount to $%
\partial P/\partial \mu >0$) in the entire existence region. For the AS
solution given by Eq. (\ref{eq37}), the calculation of norm (\ref{eq38})
yields a result which does not depend on $\varepsilon ,$ cf. Eq. (\ref%
{independent}) for the AS solution in System 1:
\begin{equation}
P=\frac{(1+s^{2})(1-s^{4})}{s(1+s^{4})},  \label{eq41}
\end{equation}%
whose VK slope,\ $\partial P/\partial \mu $, is again positive in the entire
region where this solution exists.

Lastly, for the AnS solution, with $A=-B,$ the power (\ref{eq39}) is
\begin{equation}
P=\frac{2(1+s^{2})^{2}}{1+s^{4}}\left( \frac{1-s^{2}}{s}+\varepsilon \right)
,  \label{eq43}
\end{equation}%
whose slope may change its sign. The eventual conclusion is that the AnS
solution satisfies the VK criterion only in a part of its existence region,
\textit{viz}., at
\begin{equation}
\varepsilon >{\varepsilon }_{s}^{(2)}=\frac{1+s^{2}}{4s^{3}(1-s^{2})}-\frac{%
1-s^{2}}{s},  \label{eq45}
\end{equation}%
cf. similar result (\ref{eq28}) obtained above for the AnS solutions in
System 1.

\section{Numerical results}

To verify the predictions of the VA, we numerically solved
stationary equations (\ref{eq3}) and (\ref{eq29}) for both models,
applying the relaxation method and an algorithm based on the
modified Powell minimization method \cite{nashi}. The initial
guess used to construct fundamental solitons centered at the
interface of the linked semi-infinite chains which constitute
System 1 (Fig. \ref{fig1}) was taken as $U_{0}$ $=U_{1}=A>0$
for SyS solutions, $U_{0}$ $=A>0,\,U_{1}=B>0$ for AS solutions, and $%
U_{0}=A>0,U_{1}=-A$ for ones of the AnS type, with the VA-predicted values
of $A$ and $B$, while in initial values of the lattice field are set to be
zero at all other sites.

The initial ansatz for the parallel-coupled lattices which
constitute System 2 (Fig. \ref{fig4}) was $U_{0}$ $=V_{0}$ $=A>0$
for SyS modes, $U_{0}$ $=A>0,\,V_{0}$ $=B>0$ for AS modes, $U_{0}$
$=A>0,V_{0}$ $=-A$ for modes of the AnS type, and $U_{n}=V_{n}=0$
at all other sites. The results presented
here are obtained for identical coupled chains with $N_{1}=N_{2}=50$ or $%
N_{1}=N_{2}=51$ sites, for Systems 1 and 2, respectively. In other words,
the total number of sites is $N=100$ in System 1, and $N=102$ in System 2,
which features the parallel chains. The link between the chains in System 1
was set between the sites with indexes $n=0$ and $n=1$, and in System 2 the
transverse coupling between the parallel lattices was introduced at $n=0$.

The stability of the stationary modes was first checked by dint of the
linear-stability analysis, i.e., the calculation of eigenvalues (EVs) for
modes of small perturbations, following the lines of Refs. \cite{nashi}. The
respective calculations were performed in parameter space ($\varepsilon $, $%
\mu $). Then, the results were verified in direct numerical simulations of
full equations, (\ref{eq3a}) and (\ref{eq1c2}). The simulations were based
on a numerical code which used the sixth-order Runge-Kutta algorithm, as in
Ref. \cite{nashi}. The simulations were initialized by taking the stationary
soliton profiles, to which random perturbations were added.

\begin{figure}[h]
\center\includegraphics [width=15cm]{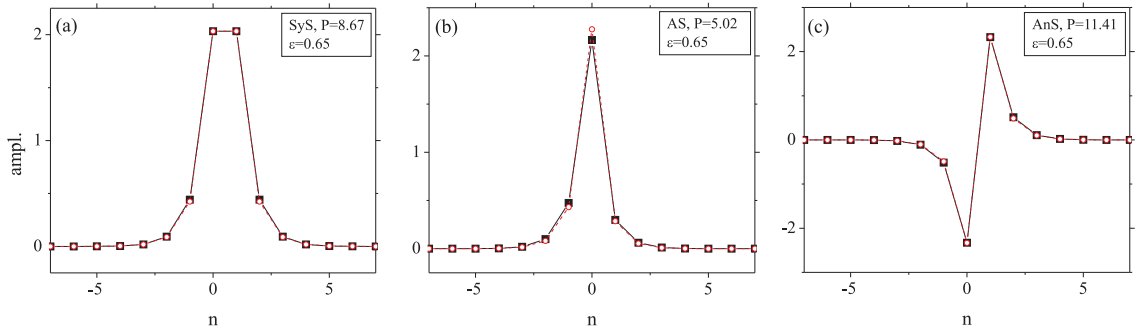} \caption{Profiles
of the fundamental solitons (interface modes) in System 1: (a)
symmetric (SyS), (b) asymmetric (AS), and (c) antisymmetric (AnS).
The inter-chain linkage strength is $\protect\varepsilon =0.65$.
The total powers corresponding to each mode are indicated in the
panels. Black solid lines with squares denote numerically
generated solitons, while red dashed lines with circles denote
variationally obtained solitons.} \label{fig7}
\end{figure}

\begin{figure}[h]
\center\includegraphics [width=15cm]{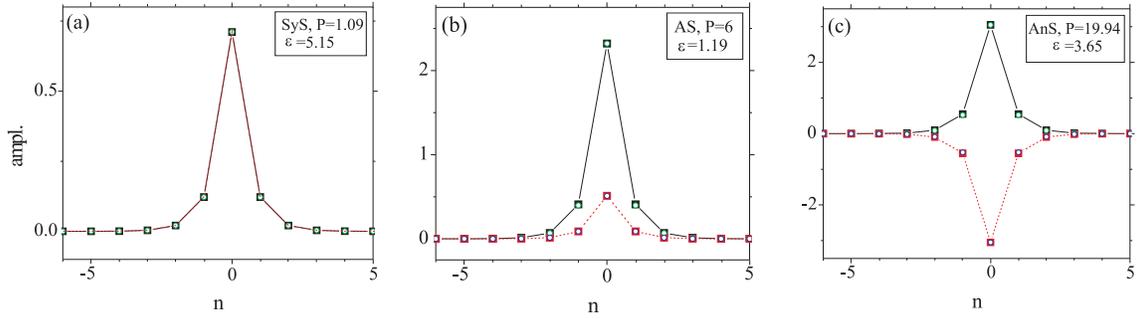} \caption{(Color
online) Profiles of fundamental solitons in System 2 (two parallel
lattices, L$_{1}$ and L$_{2}$, coupled by the transverse link with
strength $\protect\varepsilon $), obtained for $\protect\mu =6$.
The values of $\protect\varepsilon $ and the total powers are
indicated in the panels.
Black solid and red dashed lines show the profiles in lattices L$_{1}$ and L$%
_{2}$ obtained numerically. The corresponding soliton profiles predicted by
the VA are denoted by hollow squares (green for L$_{1}$ and blue for L$_{2}$%
). (a) A symmetric (SyS) soliton, (b) an asymmetric (AS) soliton,
and (c) an antisymmetric (AnS) soliton. } \label{fig8}
\end{figure}

Typical shapes of symmetric, asymmetric and antisymmetric solitons found in
the numerical form are displayed in Fig. \ref{fig7} for System 1 and in Fig. %
\ref{fig8} for System 2. In the same figures, the numerical shapes
are compared with those obtained by means of the VA. The
respective dependencies of the solitons' amplitudes $A$ and $B$ on
parameter $\varepsilon $, which accounts for the linkage between
the two chains, are displayed for all the soliton species,
alongside their VA-predicted counterparts, in the above figures
\ref{fig2} and \ref{fig3}. Further, the numerical results
demonstrate that \emph{all} the species of the solitons, including
the AnS, exist in bounded regions of the parameter space, as shown
in the above figures \ref{fig3} and \ref{fig6}, for Systems 1 and
2, respectively.

The comparison with the numerical results demonstrates that the predictions
of the VA for the symmetric and asymmetric discrete solitons are very
accurate. However, the VA fails to predict the existence borders for the
antisymmetric modes. A reason for the latter problem may be that the strong
interaction force at the interface, in the case of the opposite signs of the
lattice fields at adjacent sites, see Figs. \ref{fig7}(c) and \ref{fig8}%
(c), makes the simple form of the VA adopted above inaccurate.

In the instability region, the SyS solutions are characterized by the EV
spectrum consisting of real pairs. Under small asymmetric perturbations, an
unstable SyS sheds off a part of its total power and relaxes into an
asymmetric breather with a smaller power, as shown for System 1 in Fig. \ref%
{fig9} (a). As in other conservative systems, \cite{Akhmediev}-\cite{2D-BEC}%
, \cite{VA}, the asymmetric breathers cannot readily self-trap into
stationary solitons. In System 2, the behavior of the unstable symmetric
solitons is quite similar.

\begin{figure}[h]
\center\includegraphics [width=13cm]{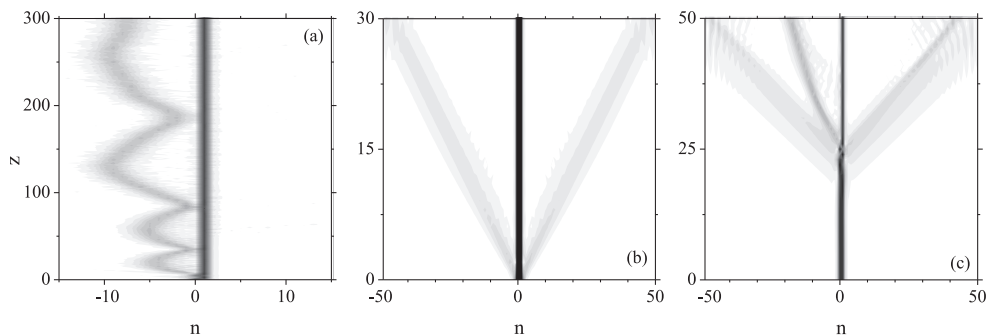} \caption{Typical
examples of the evolution of perturbed unstable solitons in
System 1. (a) A symmetric soliton, with $\protect\varepsilon =0.65,\,\protect%
\mu =6.425$. Other panels display antisymmetric solitons, with $\protect\mu %
=6$: (b) $\protect\varepsilon =9.1$ (real unstable EVs, see Fig. \protect\ref%
{fig5}), and (c) $\protect\varepsilon =1.15$ (complex EVs). The unstable
symmetric soliton, after shedding off a part of its power, evolves into an
asymmetric breather. The unstable antisymmetric solitons transform into
breathers too, see the text.}
\label{fig9}
\end{figure}

In both Systems 1 and 2, the EV spectra for the stationary asymmetric modes
are stable in the entire existence region. This is also confirmed by direct
simulations, which show that perturbed AS develop small-amplitude persistent
oscillations without any trend to destruction.

The antisymmetric solitons change their stability twice, in both
configurations considered, 1 and 2. With the decrease of
$\varepsilon $ at fixed $\mu $, the unstable AnS branch, which is
characterized by a pair of real EVs [in the area between the green
dotted and black solid lines in Figs. \ref{fig3} (b) and
\ref{fig6} (b)], changes into the stable one. With the further
decrease of $\varepsilon ,$ the AnS loses its stability through
the appearance of two pairs (a quartet) of complex EVs with
significant real parts [in the area below the red dashed line in
Figs. \ref{fig3} (b) and \ref{fig6} (b)]. For example, the
antisymmetric modes corresponding to $\mu =5$ acquire the
oscillatory instability, with the decrease of $\varepsilon $, at
$\varepsilon \approx 1.5$ and $2$ in Systems 1 and 2,
respectively, which is indicated by the appearance of a quartet of
complex EVs in their spectra (in the areas below the red line in
Figs \ref{fig3} and \ref{fig6}). Note that the VA has also shown
that the AnS change their stability according to the VK criterion,
see Eqs. (\ref{eq28})
and (\ref{eq45}). However, the corresponding VA-predicted critical curves, $%
\varepsilon _{c}^{(1,2)}(\mu )$, are found to be situated beyond the
numerically obtained existence regions of the AnS solutions.

The predictions concerning the stability of the AnS were checked
by direct simulations. In Fig. \ref{fig9}(b) and (c), the
evolution of typical unstable antisymmetric modes in System 1 is
displayed. Panels (b) and (c) in this figure illustrate the
evolution of the unstable AnS whose EV spectrum contains a pair of
real EVs, or a complex EV quartet, respectively. The unstable
antisymmetric modes radiate away a part of their power, relaxing
to antisymmetric (b) or asymmetric (c) interface modes, which
exhibit small-amplitude persistent oscillations. In System 2, the
same happens with the unstable AnS belonging to the instability
area below the red (dashed) line in Fig. \ref{fig6}. However, in
contrast to that, the AnS in System 2 which is characterized by
the EV spectrum with a pair of real EVs turn out to be
\emph{robust} under the action of small perturbations. They do not
emit radiation waves, staying localized and strongly pinned to the
link connecting the two infinite chains.

Returning to the global existence diagrams, it is worthy to note that two
\textit{bistability areas} can be identified in both systems, 1 and 2: the
domain of the coexistence of stable symmetric and antisymmetric solitons, or
the one featuring the simultaneous stability of asymmetric and antisymmetric
modes, on the opposite sides on the SSB bifurcation. This result is in
accordance with similar findings reported in other linearly-coupled
two-component systems featuring the self-focusing nonlinearity \cite%
{1D-BEC,2D-BEC}.

\section{Conclusion}

In this paper, we have investigated several species of fundamental
localized modes formed at the interface between two linearly
coupled lattices (chains) with the cubic on-site nonlinearity. Two
configurations were considered: the one with the linkage between
two semi-infinite chains (in other words, the usual discrete NLS
model with a spring defect), and two infinite chains placed in
parallel planes which are coupled by the transverse link at one
site. These systems can be readily implemented as arrays of
optical waveguides.

In both models, the VA\ (variational approximation) was used to predict the
existence regions, in the parameter space of $(\mu ,\varepsilon )$ (the
propagation constant and the strength of the coupling between the two
subsystems), for the localized symmetric, asymmetric and antisymmetric
discrete solitons pinned to the interface. The stability of the modes was
predicted as per the VK criterion and general properties of the
supercritical pitchfork bifurcation, which destabilizes the SyS (symmetric
solitons) giving birth to AS (asymmetric solitons). The predictions were
verified against numerical results, as concerns the existence and stability
of all the soliton species. In both systems considered, the existence
regions of all the localized modes are bounded. The AS are stable in their
entire existence domain. The existence and stability domains for AnS
(antisymmetric solitons) were found in the numerical form. Both systems give
rise to the bistability between the AnS on the one hand, and either SyS or
AS, on the other. Direct simulations demonstrate that unstable SyS are
transformed into asymmetric breathers. Those antisymmetric modes which are
unstable shed off a part of the total power and also evolve into breathers.
In System 2, in a part of the parametric domain where the computation of the
eigenvalues predicts the instability of the antisymmetric localized modes,
they actually evolve into strongly pinned robust spikes.

\acknowledgments Lj. H., G. G., and A.M. acknowledge support from the
Ministry of Science, Serbia (Project 141034). The work of B.A.M. was
supported, in a part, by grant No. 149/2006 from the German-Israel
Foundation.

\end{document}